\documentclass[traditabstract]{aa}

\usepackage{graphicx}
\usepackage{graphicx,psfig}
\usepackage{txfonts}
\begin{document}
\title{Bright optical day-side emission from extrasolar planet CoRoT-2b}

\author{I.A.G. Snellen \inst{1}, E.J.W. de Mooij \inst{1}, \& A. Burrows \inst{2}}

\institute{Leiden Observatory, Leiden University, Postbus 9513, 2300 RA Leiden, The Netherlands\\
\and
Department of Astrophysical Sciences, Princeton University, Princeton NJ08544, USA}

\date{}

\abstract{We present our analysis of the red-channel CoRoT data of extrasolar planet CoRoT-2b. A deep secondary eclipse is detected at a level of $1.02 \pm 0.20 \times 10^{-4}$, which suggests that all of the planet-signal detected previously in white light by Alonso et al. (2009) originates from the red channel. CoRoT-2b is the coolest exoplanet that has been detected in the optical so far. In contrast to the other planets, its measured brightness temperature of $2170\pm55$ K is significantly higher than its maximum hemisphere-averaged effective day-side temperature. However, it is not expected that a hot Jupiter radiates as a black body, and its thermal spectrum can deviate significantly from a Planck curve. We present models of the planet/star flux ratio as function of wavelength, which are calculated for a T/P profile in radiative and hydrostatic equilibrium, using a self-consistent atmosphere code. These are compared with the CoRoT detection, and with measurements at 4.5 and 8 $\mu$m from the Spitzer Space Telescope from Gillon et al. (2009). We estimate that reflected light contributes only at a 10$-$20\% level to the total optical eclipse depth. The models allow for an 'extra absorber' to be inserted at high altitude in the planet's atmosphere. This produces a thermal inversion layer, recently invoked to explain the photometric reversals and flux enhancements seen in some planets in the infrared. 
In the 0.5$-$1.5 $\mu$m wavelength range, the model-spectra of planets {\sl with} an extra absorber at high altitude, are relatively suppressed in flux compared to those {\sl without} such absorber. We therefore argue that, in contrast to the other exoplanets detected in the optical so far, CoRoT-2b may not exhibit a significant thermal inversion in its atmosphere, causing its optical brightness temperature to be  boosted above its maximum effective day-side temperature.}

\maketitle

\section{Introduction}

Hot Jupiters, a class of extrasolar planets that orbit their host star at only a few percent of the Earth-Sun distance, are principle targets for studying optical light, which can be a combination of reflected star-light and/or thermal emission. So far the optical reflectivity of these strongly irradiated planets have been found to be very low, with stringent upper limits to their geometric albedo of A$_g<$0.12 for HD75289b, and A$_g<$0.12 for HD209458b, compared to A$_g$ = 0.41 $-$ 0.52 for the solar system gas giants (Cox 2000). This is as expected from theoretical predictions, in particular in the absence of clouds in the planet atmospheres. Recently, optical emission from three extrasolar planets has been measured, by detection of the planetary eclipse of OGLE-TR-56b (Sing \& Lopez-Morales 2009), and by detection of the phase variations of CoRoT-1b (Snellen, de Mooij, \& Albrecht 2009, see also Alonso et al. 2009), and HATP-7b (Borucki et al 2009). The detected optical radiation from these planets is most likely from the short-wavelength tail of their thermal emission, and not reflected star-light, which shifts into the optical wavelength regime for these very hot planets. 

In this paper we report on our analysis of the red channel CoRoT data on the secondary eclipse of extrasolar planet CoRoT-2b, and investigate whether the deep eclipse is due to only thermal radiation or also reflected star-light. 
CoRoT-2b is an object with 3.3 Jupiter masses and 1.465 Jupiter radii, orbiting a G dwarf star (T$_{\rm{eff}}$ = 5625 K) with an orbital  period of 1.743 days (Alonso et al. 2009).  Notably, the host star shows a profound low frequency modulation in brightness at a level of 1-2\%, due to the presence of spots on its surface, exhibiting a rotational period of 4.52 days, indicating it is a young system possibly $<$500 Myr old (Lanza et al. 2009). In Section 2 the CoRoT observations, data reduction and analysis are presented. In Section 3 the red channel eclipse measurement and the 4 and 8 $\mu$ Spitzer observations (Gillon et al. 2009) are compared with models of the planet/star flux ratio as function of wavelength, produced with a self-consistent atmosphere code and solution techniques as described earlier in Burrows et al. (2008). The results are discussed in Section 4.

\section{Observations and analyses}

\subsection{CoRoT observations}

The first long run of the CoRoT satellite (Auvergne et al. 2009), from May until October 2007, was pointed towards the galactic center direction, monitoring the flux of $\sim$12,000 stars.  CoRoT-2b was identified as a planetary candidate by the alarm mode within a few days (Alonso et al. 2009). This alarm mode is aimed at identifying a reduced number of stars for which the sampling rate is changed from 512 s to 32 s. In this way, CoRoT-2b was observed almost continuously for 142 days, with a sampling rate of 512 sec for the first 5 days, and 32 sec for the remainder of the time, resulting in $\sim$369,700 flux measurements. For our analysis we used the updated N2 pipeline data\footnote{http://idc-corotn2-public.ias.u-psud.fr/}, which was released to the general public on February 16, 2009. 

A prism in front of the exoplanet-CCDs produces a small spectrum for each star, on which aperture photometry is performed in three bands (red, green, and blue; Auvergne et al. 2009) This is done on board to comply with the available telemetry volume. The transmission curves of the three bands are different for each targeted star, and depend on the template chosen for the on-board aperture photometry, which in itself is based on the effective temperature of the star and its position on the CCD. For CoRoT-2 template ID number 25 was used with the red channel covered by 38 pixels. The background was measured on 400 windows of 10$\times$10 pixels in the field, of which 300 and 100 with sampling of 512 s and 32 s sampling respectively (Auvergne et al. 2009). 

To determine the short wavelength cutoff of the red channel, we first compared the Kurucz model spectrum (Kurucz 1993) for the host star with its apparent magnitude in B, V, r, I, J, H, and K bands as taken from the EXODAT and 2MASS databases (Meunier et al. 2007; Skrutskie et al. 2006). We found that the optical to near-infrared colors are best fitted with an interstellar extinction of A$_{\rm{V}}$=0.7. A faint star at 4 arcseconds from CoRoT-2 falls completely within its mask template. From its broadband colors and apparent brightness we determine it to be an M-dwarf with an effective surface temperature of T$\sim$3500 K, 3.5 magnitudes fainter than CoRoT-2 in V-band. The short wavelength cut-off of the red channel pass-band was subsequently determined from the overall transmission curve for the telescope/CCD combination multiplied with the appropriately reddened Kurucz model spectrum of the host star plus a small contribution from the M-dwarf, which was compared to the fraction of photons collected in the red channel. This results in a wavelength cutoff of 580 nm, and an effective wavelength of 720 nm.
\subsection{CoRoT data analysis}

The median red-channel flux of the star is $\sim$503,000 e- per 32 s exposure, with an average background level of 18,700 e$^{-}$, resulting in a photon noise limit of 0.00144. To avoid systematic effects originating from the beginning or end of the light curve, or from the change in sampling, we only used the 32 sec sampled data from the first to the final transit, covering a total of $\sim$132 days and 78 planetary orbits. Only those flux measurements which are assigned status = 0 were included in the analysis (90\% of the remaining data), indicating that they are a valid measurement, and avoiding those influenced by an energetic particle hit or glitch, or those acquired during the crossing of the satellite over the South Atlantic Anomaly. We further rejected 3$\sigma$ outlier points relative to the median-smoothed light curve (1.4\% of the data).

\begin{table}

\caption{The star and planet parameters from Alonso et al. (2008), which are used for our analysis. Parameter P is the orbital period, T$_c$ is the transit time, T$_*$ is the effective temperature of the star, M$_*$ and R$_*$ are the stellar mass and radius, R$_p$/R$_*$ is the ratio of planet to star radius, e is the eccentricity, and I is the orbital inclination.}

\begin{tabular}{ll}\hline
&\\
P[d]      &= 1.7429964$\pm$0.0000017\\
T$_c$[d]     &= 2454237.53562$\pm$0.00014\\
T$_*$ [K]    &= 5625$\pm$120\\
M$_*$ [M$_{\rm{sun}}$] &= 0.97$\pm$0.06\\
R$_*$[R$_{\rm{sun}}$]  &= 0.902$\pm$0.018\\
R$_p$/R$_*$     &= 0.1667$\pm$0.0006\\
e         &= 0 (fixed)\\
I [deg]   &= 87.84$\pm$0.10\\
&\\\hline
\end{tabular}
\end{table}

\begin{figure}
\psfig{figure=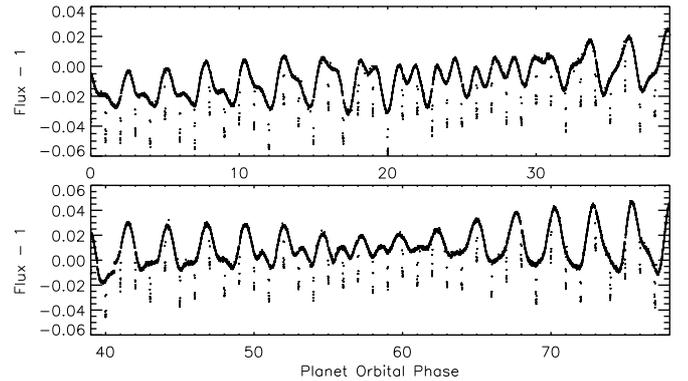,width=0.5\textwidth}
\caption{The red-channel light curve covering 78 planet orbital periods of CoRoT-2b, as used for our analysis. The data have been binned to 0.005 in phase. Intrinsic stellar variability, probably due to spots, completely dominates variability on time scales $>$ 1 day (see Lanza et al. 2009). The point-to-point scatter, relative to the median-smoothed light curve is 5.5$\times$10$^{-4}$.}
\end{figure}

\begin{figure}
\psfig{figure=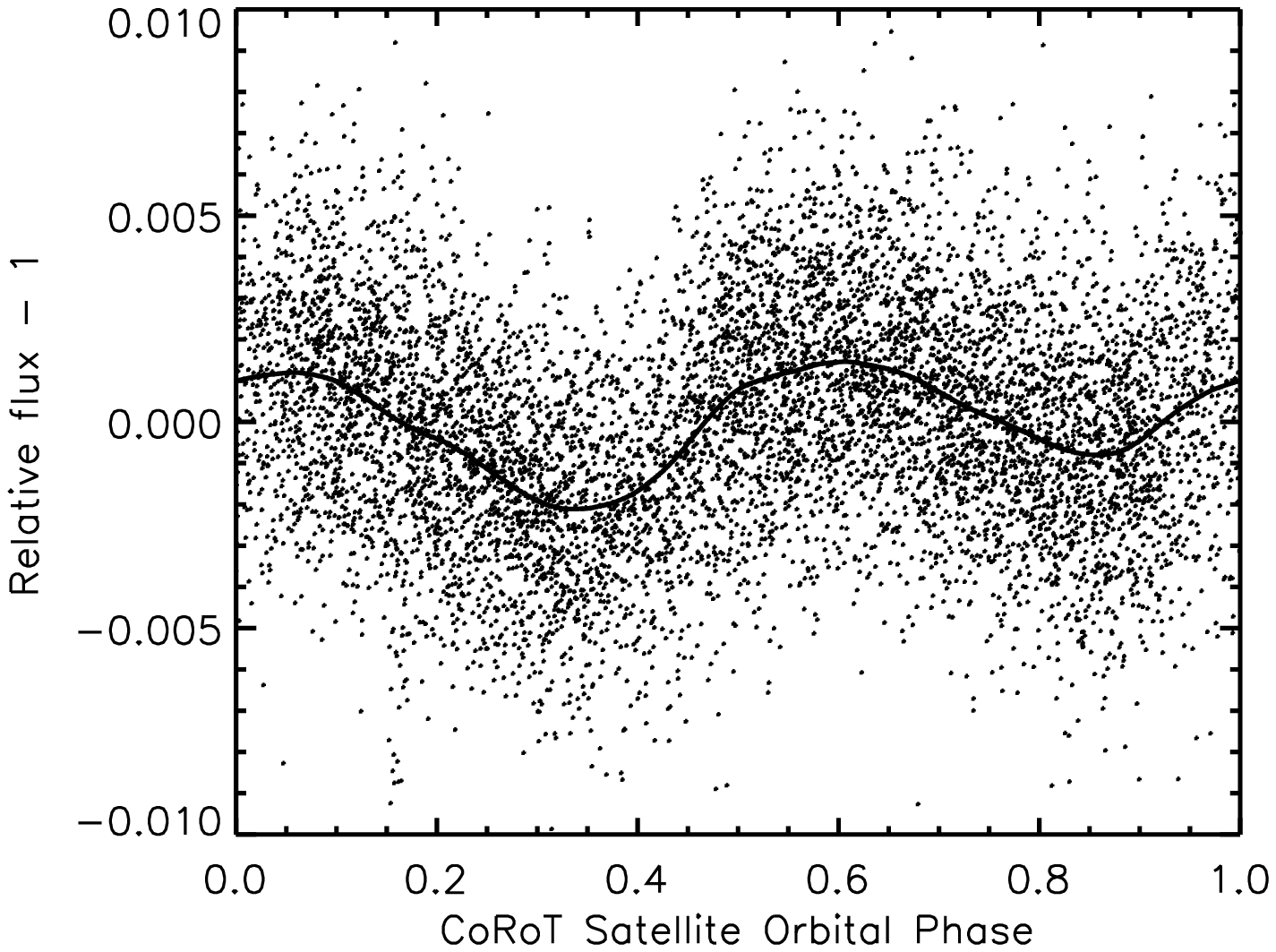,width=0.45\textwidth}
\psfig{figure=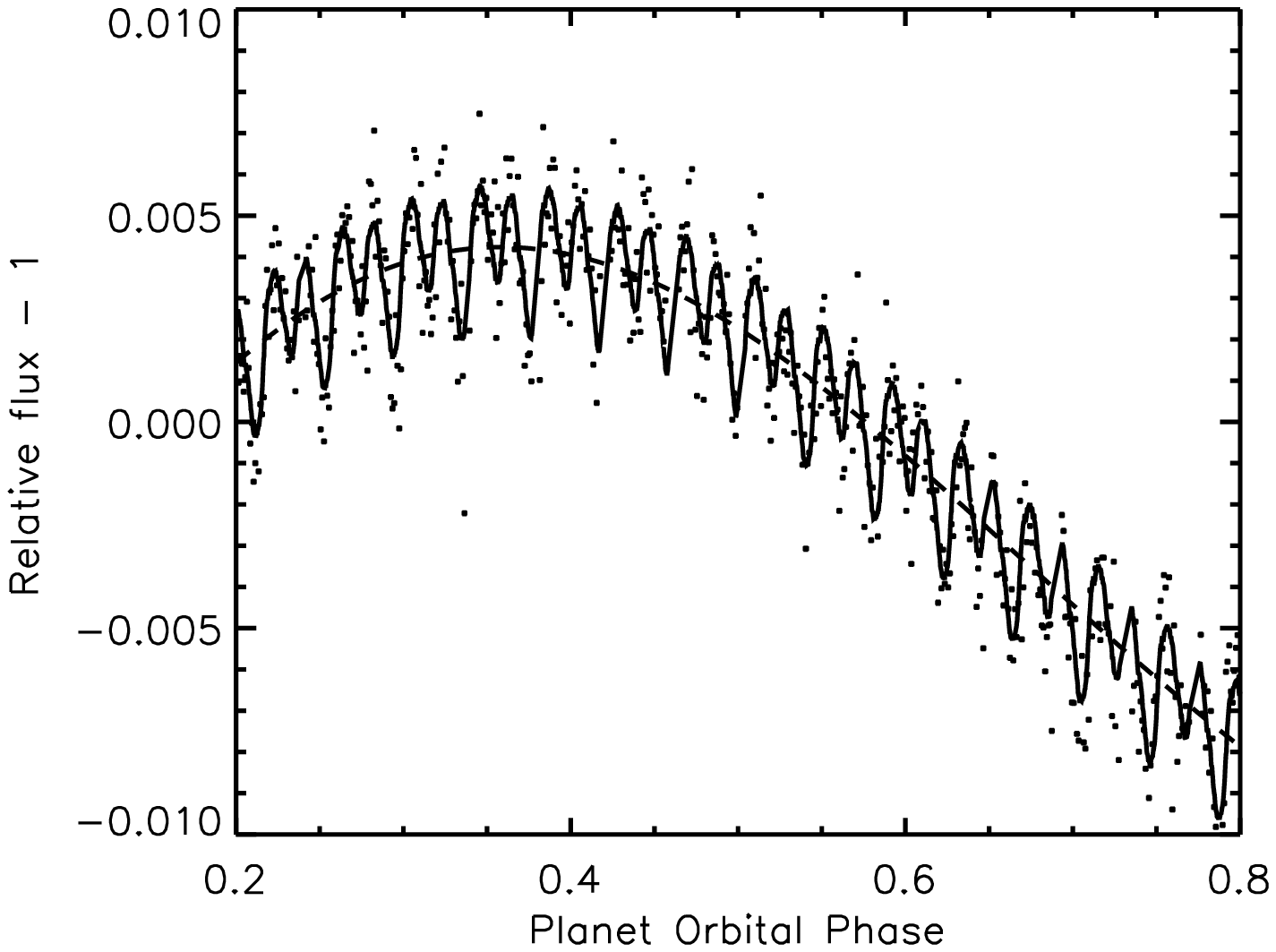,width=0.45\textwidth}
\caption{[top panel] Data covering two planetary orbits, phase-folded over the orbital period of the CoRoT Satellite (103 min), showing the instrumental jitter at a level of $\sim$0.002. The solid line indicates the correction for this effect for this section of the data. [lower panel] Small part of the light curve showing the same 103-min jitter and its correction (solid line). The dashed line is our fit to the long-timescale stellar variability which was first removed before the instrumental correction was determined, after which it was added again.  }
\end{figure}

\begin{figure}
\psfig{figure=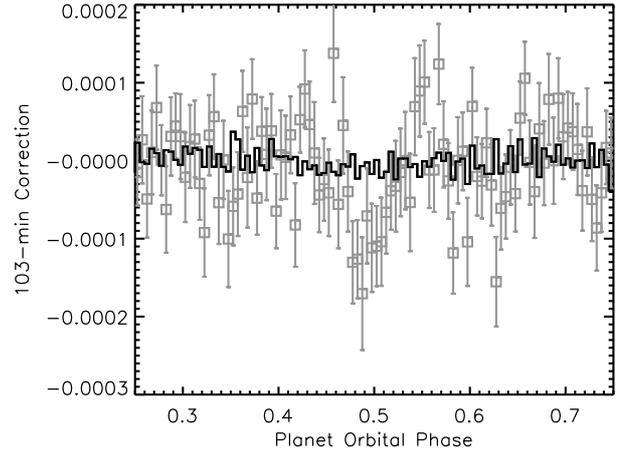,width=0.45\textwidth}
\caption{The overall correction (solid line) for the 103 min perturbation to the final binned light-curve (grey squares with 1σ error bars). Although the corrections have an amplitude at a level of 0.002 (more than an order of magnitude larger than the measured eclipse depth), they largely cancel out over the $\sim$1900 satellite orbits. Assuming an uncertainty of 10\% in the amplitude of the 103-min perturbation signal, it contributes to the errors in the final light curve at an insignificant level of 2$\times$10$^{-6}$ per data point.}
\end{figure}

The Earth has a significant influence on the photometric performance of the satellite and introduces relevant perturbations on time scales of the satellite orbital period (103 min) and the 24 hr day, due to ingress and egress of the spacecraft from the Earth’ shadow, variations in the gravity field and magnetic field (such as the South Atlantic Anomaly), and changes in the levels of thermal and reflected light from the Earth. Most of these effects were corrected for prior to the data release for the general astronomy community (Auvergne et al. 2009), however, residual effects are still present in the data.  The variations in the light curve of CoRoT-2 are dominated by intrinsic stellar variability at a level of 1-2\%, on time scales of a few days (see Fig. 1). The residual 24 h instrumental perturbations, as found in the data of CoRoT-1, are therefore indistinguishable from the stellar variability and cannot be corrected for, but we do correct for the perturbations on a 103-min time scale.

We corrected for this 103 min perturbation, which is associated with the satellite orbital period, as illustrated in Figure 2. First we determined the average flux of CoRoT-2 per 103 min. period, and constructed a model of the long-term variability by fitting a 7th order polynomial through these points (excluding those affected by transits), for blocks of two planet orbital periods (corresponding to $\sim$5.4 days). This model was used to normalize the light curve. We subsequently phase-folded the normalized light-curve of these blocks over 103 minutes and box-car smoothed over 5 minutes. These curves were used to correct each associated block of data. In this way, the relative standard deviation was reduced from 0.0067 to 0.0033, still more than a factor of two above the photon noise limit. Afterwards, the long-term variability in the data was restored by multiplying the light curve with the initial model. Hence, in total we determined and corrected for the 103-min perturbation for 39 blocks of data, each covering 2 planetary orbital periods. In Figure 3 we show the total influence of these corrections on the phase-folded data. Assuming an uncertainty of 10\% on the amplitude of the corrections, they contribute to the errors in the final light curve at an insignificant level of 2$\times$10$^{-6}$ per data point.

\subsection{Binning, fitting, and phase-folding}

To reduce the noise per data point and expose possible systematic effects, we averaged and binned the data over regular phase steps of $\phi$ = 0.005, corresponding to 12.5 minutes. The relative standard deviation of this binned data ($\sim$15,000 points) relative to a median-smoothed light curve is 0.00055 (compared to 0.00031 as expected from the photon noise limit). It exposes several positions in the light curve which show sudden jumps in flux (Fig. 4). When these occur at a relevant section of the planetary phase (0.25$<$$\phi$$<$0.75), the data taken during those particular planetary orbits were not used for further analysis. We removed data from 6 of the 78 orbits in this way.   To remove the influence of the long-term intrinsic variability of the star, we least-squares fitted the data for each planet-orbit with a third order polynomial between phase 0.25$<$$\phi$$<$0.45 and 0.55$<$$\phi$$<$0.75, making sure that the fit is not influence by the possible presence of a secondary eclipse at $\phi$=0.5. The best-fit coefficients were subsequently used to correct for the stellar variability, by constructing a polynomial curve for the same range in phase including the possible eclipse, 0.25$<$$\phi$$<$0.75.  To check for systematic effects, we varied the range in phase and order of the polynomial to fit the long-term variability (see below). Finally, the data from all 72 planet orbits were averaged in phase, and their associated errors were estimated from the variance between the different orbits, typically $\sim$6$\times$10$^{-5}$ (compared to 3.7$\times$10$^{-5}$ as expected from the photon noise limit).

\begin{figure}
\vspace{1cm}
\psfig{figure=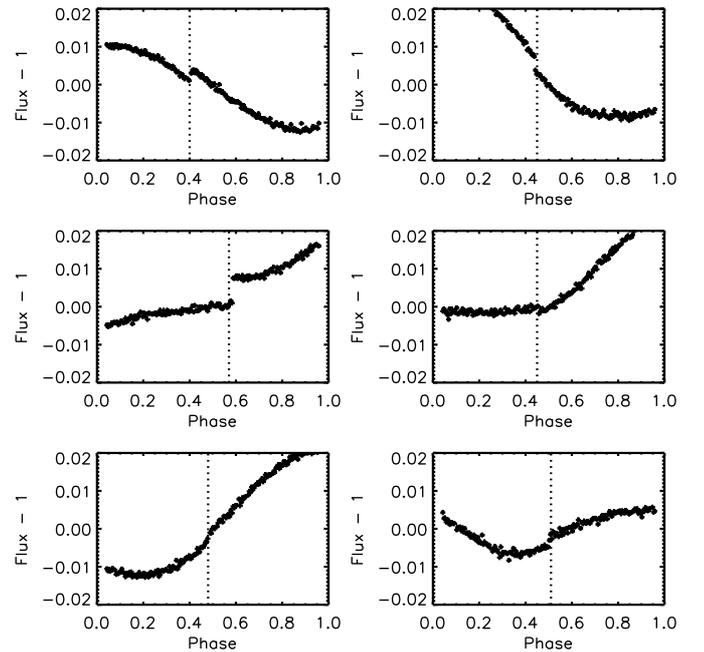,width=0.48\textwidth}
\caption{Six segments of the light curve, each one planet orbital period in length, showing sudden jumps in flux. The data from these planet orbits were omitted for further analysis.}
\end{figure}

\subsection{Chi-squared analysis of the eclipse}

\begin{figure}
\psfig{figure=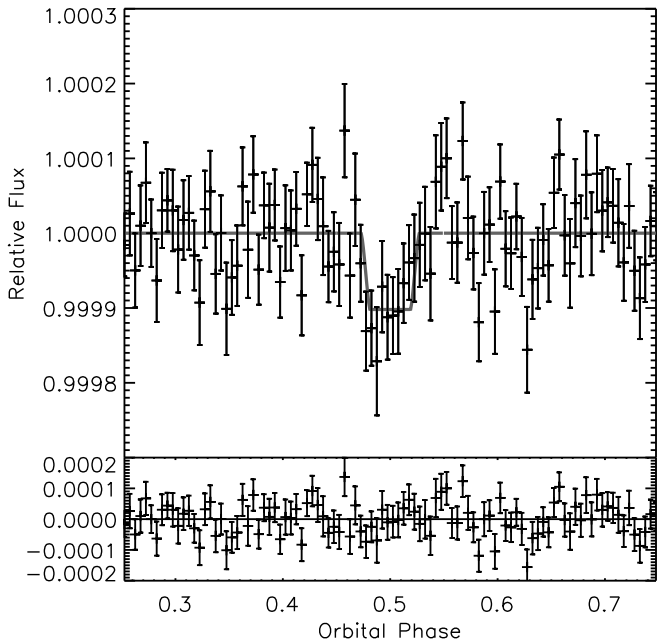,width=0.5\textwidth}
\caption{The secondary eclipse of CoRoT-2b. The phase folded light-curve is shown from 142 days of CoRoT monitoring in its red channel. The data are binned over 0.005 in phase. The integrated flux from the day-side of the planet relative to the stellar flux, as determined from the depth of the eclipse, is found to be 1.02$\pm$0.20$\times$10$^{-4}$. This corresponds to a brightness temperature of T$_{\rm{B}}$=2170$\pm$55 K, which is significantly higher than the maximum possible effective day-side temperature for CoRoT-2b.}
\end{figure}

The final phase-folded light curve (Fig. 5) shows a distinct dip at $\phi$=0.5, exhibiting the correct width to be the planet's secondary eclipse. 
It was fitted with a three parameter model using a chi-squared analysis, with K, the eclipse depth, $\phi_0$, the eclipse timing, and Z$_0$, the normalization outside the eclipse, as free parameters. The shape of the secondary eclipse as function of phase is calculated from the planet and star parameters (Table 1) using the algorithm of Mandel and Agol (2002). The analysis was performed on a grid in 3-dimensional parameter space, for which the constraints on a single parameter were determined by marginalizing over the others. The errors in the data points were increased by 14\% to obtain a reduced chi-squared of unity. In this way, we determined the two relevant parameters to be K=1.09$\pm 0.19 \times 10^{-4}$ and $\phi_0=0.495 \pm 0.003$. Since the deviation of the eclipse timing from $\phi$=0.5 is not significant, and $\phi_0$=0.5 is the likely solution for such a close-in planet, leaving the eclipse timing to vary freely the chi-squared analysis may bias K towards higher values. If the eclipse timing is kept at $\phi_0$=0.5, the eclipse depth is measured to be K =1.02$\pm 0.20 \times10^{-4}$, which we use as our final result. Note that Gillon et al. (2009) measure a deviation from a circular orbit using Spitzer, but at a smaller level, with e$cos$$\omega$=$-$0.0029$\pm0.0006$, which corresponds to  $\phi_0=0.4982 \pm 0.0004$, $\sim$1$\sigma$ away from our measurement.

\subsection{Markov-chain Monte Carlo Simulation}

As an extra check for a degeneracy between the parameters, we also performed a Markov-Chain Monte Carlo (MCMC) simulation (Tegmark et al. 2004), using the Metropolis-Hastings algorithm (Metropolis et al. 1953; Hastings 1970), which should result in similar uncertainties in the model parameters as above, since the full 3-dimensional parameter space was already mapped. Five independent chains were created, each from a random initial position, consisting of 50,000 points, for which the first 10\% of each chain were discarded to minimize the effect of the initial condition. The Gelman \& Rubin statistic (Gelman \& Rubin 1992) was calculated for each parameter to check the convergence and the consistency between the chains, which is a comparison between the intra-chain and inter-chain variance. The results were well within 1\% of unity, a sign of good mixing and convergence. The estimated probability distribution of K is shown in the middle panel of Figure 6, with the median of the distribution indicated by a solid line. The dashed lines enclose 68\% of the results, with equal probability on either side of the median. This corresponds to K = 1.04$\pm 0.22 \times 10^{-4}$. The probability distribution of $\phi_0$ is shown in the top panel of Figure 6. It corresponds to $\phi_0$=0.495$\pm$0.003. The lower panel of Fig. 6 displays $\phi_0$ versus K for the MCMC simulation, showing a randomly selected subsample of 2$\times$10$^{3}$ points for plotting purposes.

\begin{figure}
\psfig{figure=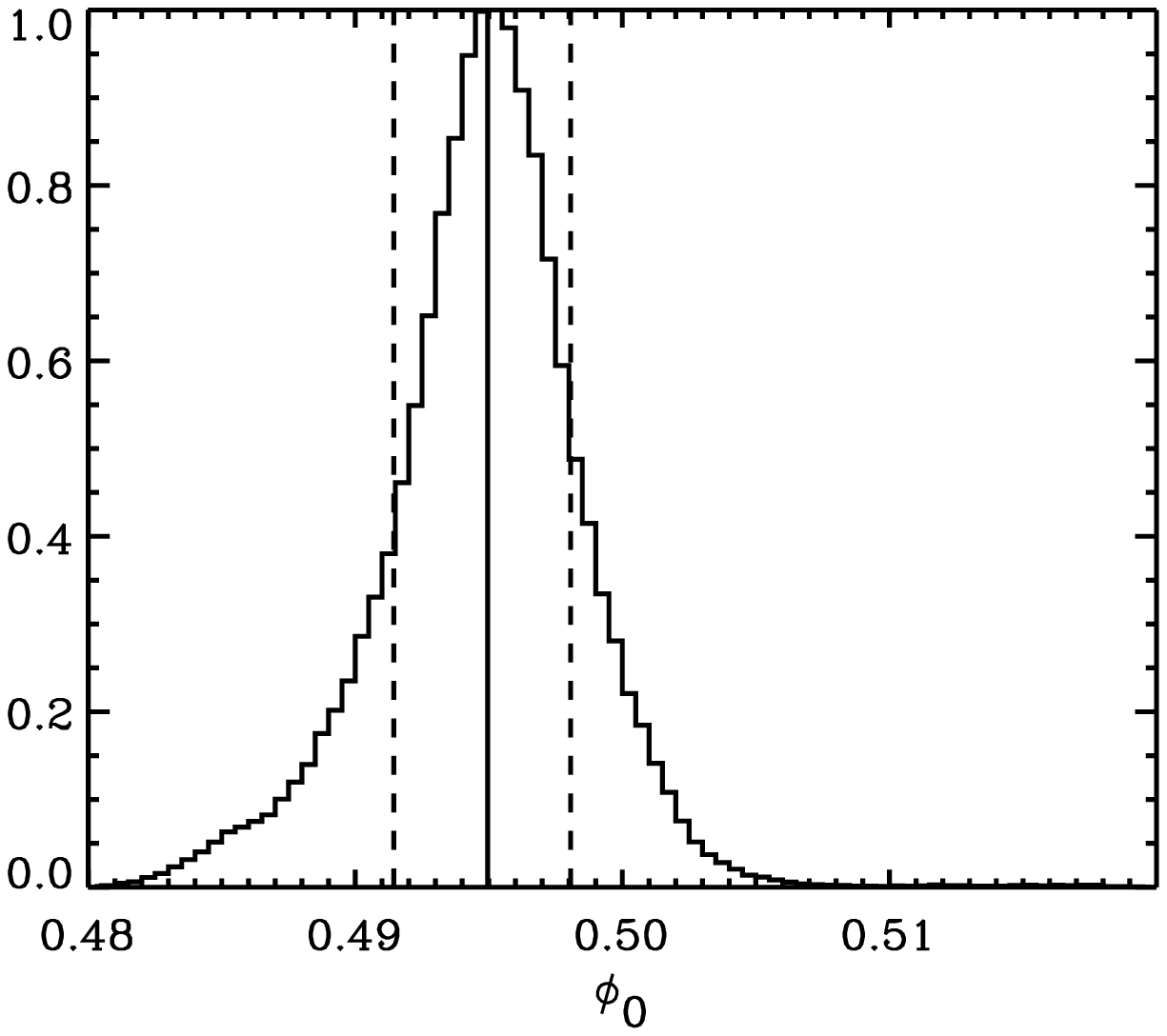,width=0.4\textwidth}
\psfig{figure=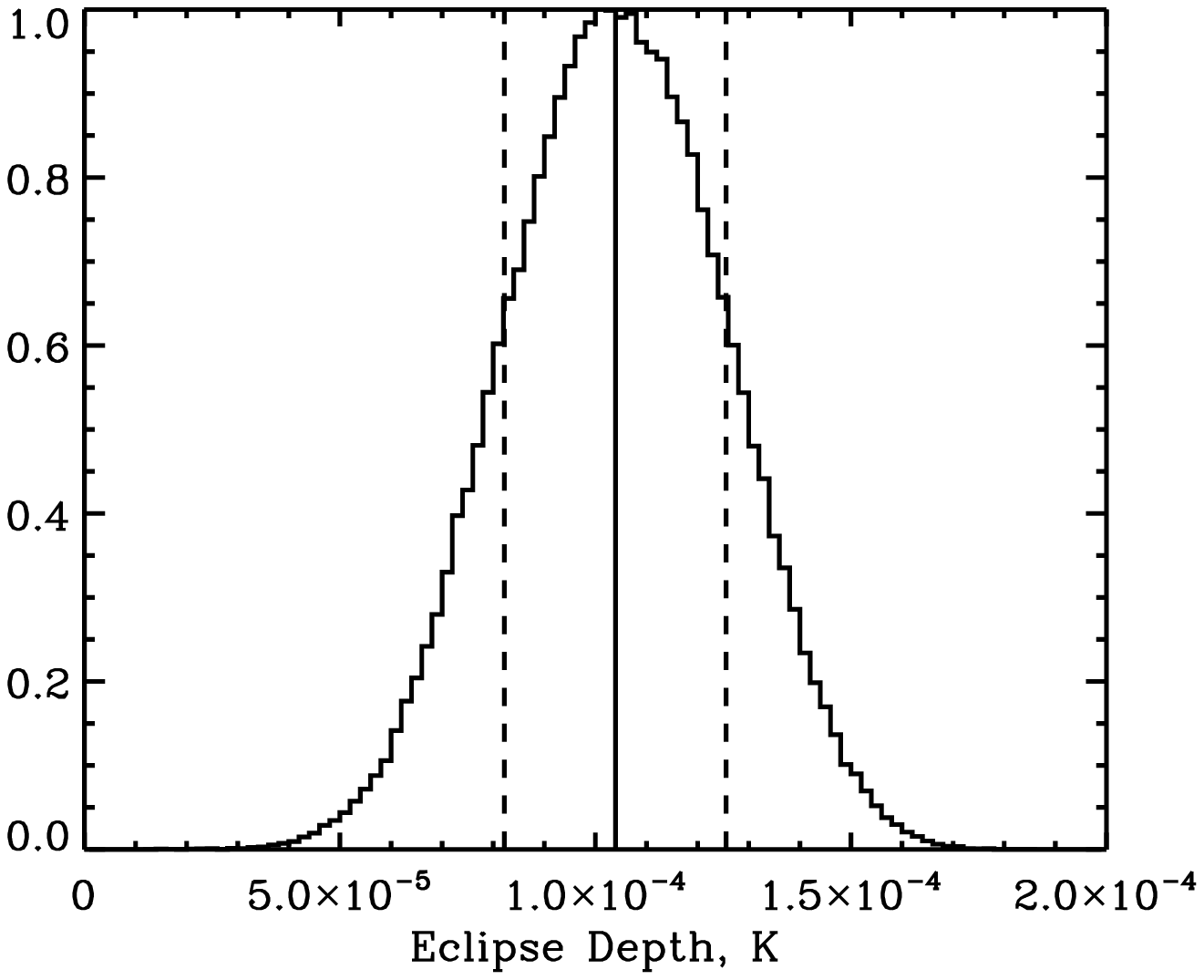,width=0.4\textwidth}
\psfig{figure=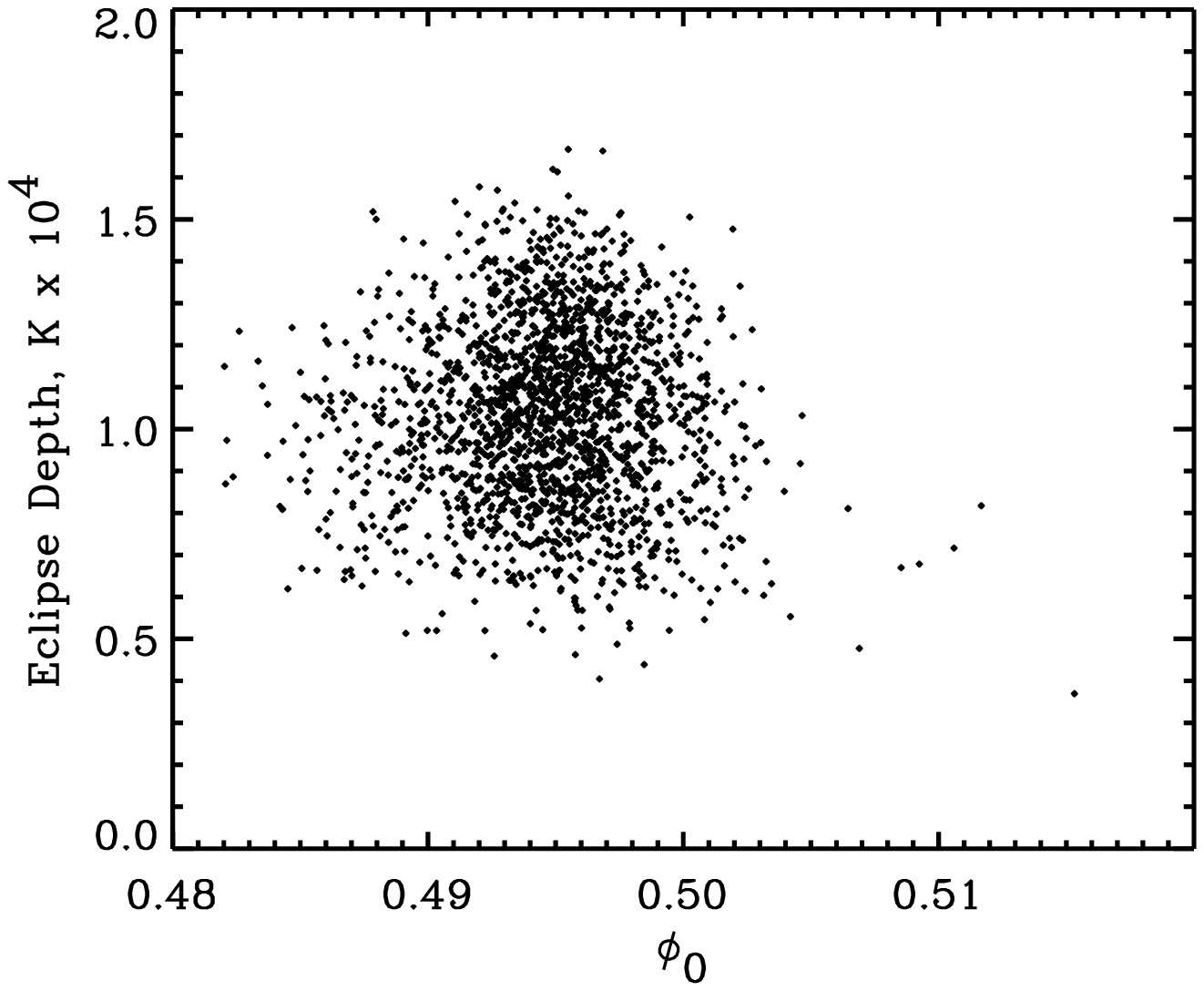,width=0.4\textwidth}
\caption{The probability distributions of $\phi_0$ (top panel) and the eclipse depth, K (middle panel) from the MCMC simulations. The solid line shows the median value. The dashed lines enclose 68\% of the results with equal probability on either side of the median. The lower panel shows  $\phi_0$ versus K from the MCMC simulations, showing a randomly selected
subsample of 2$\times$10$^{3}$ points for plotting purposes.}
\end{figure}

\begin{figure}
\psfig{figure=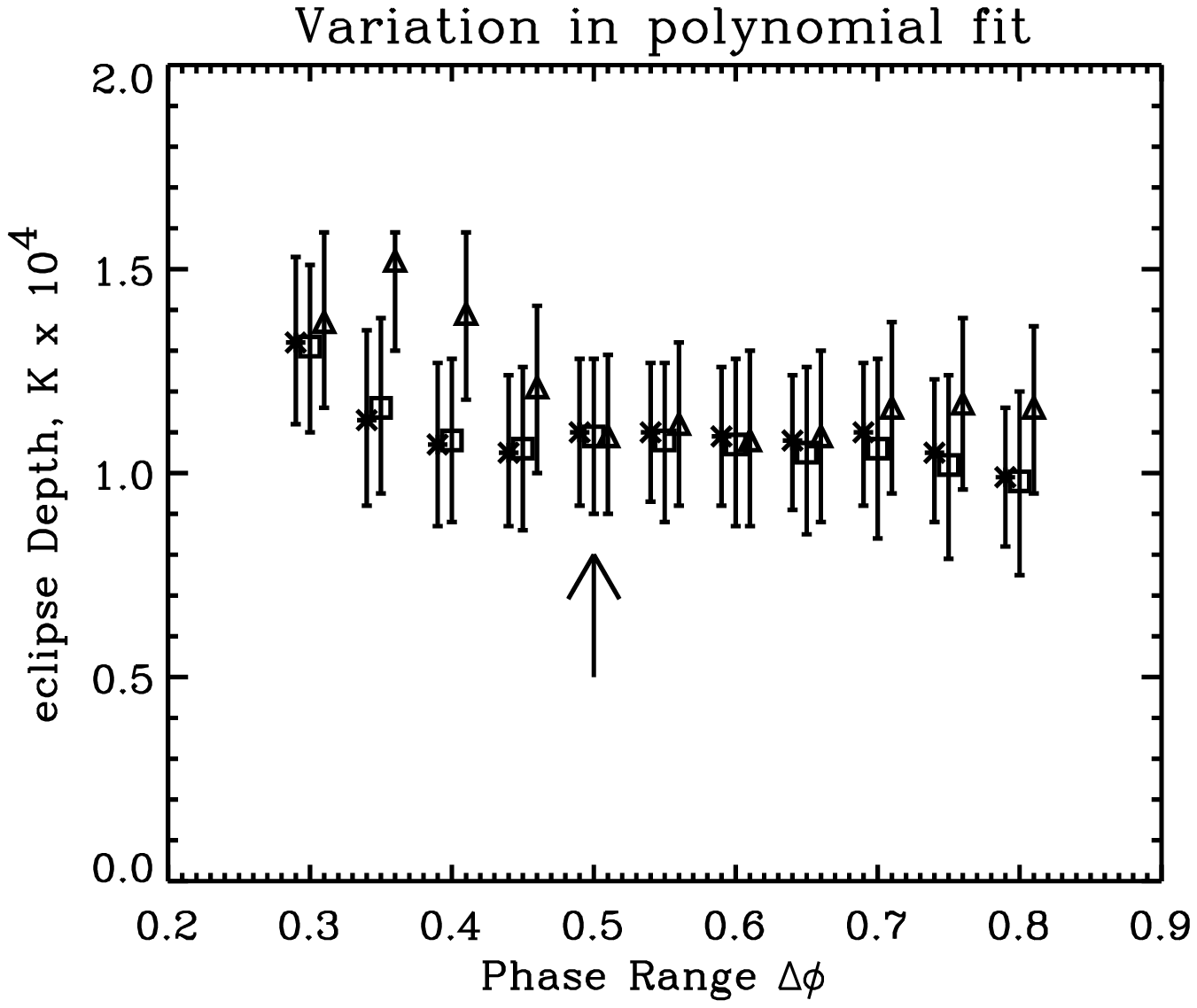,width=0.45\textwidth}
\psfig{figure=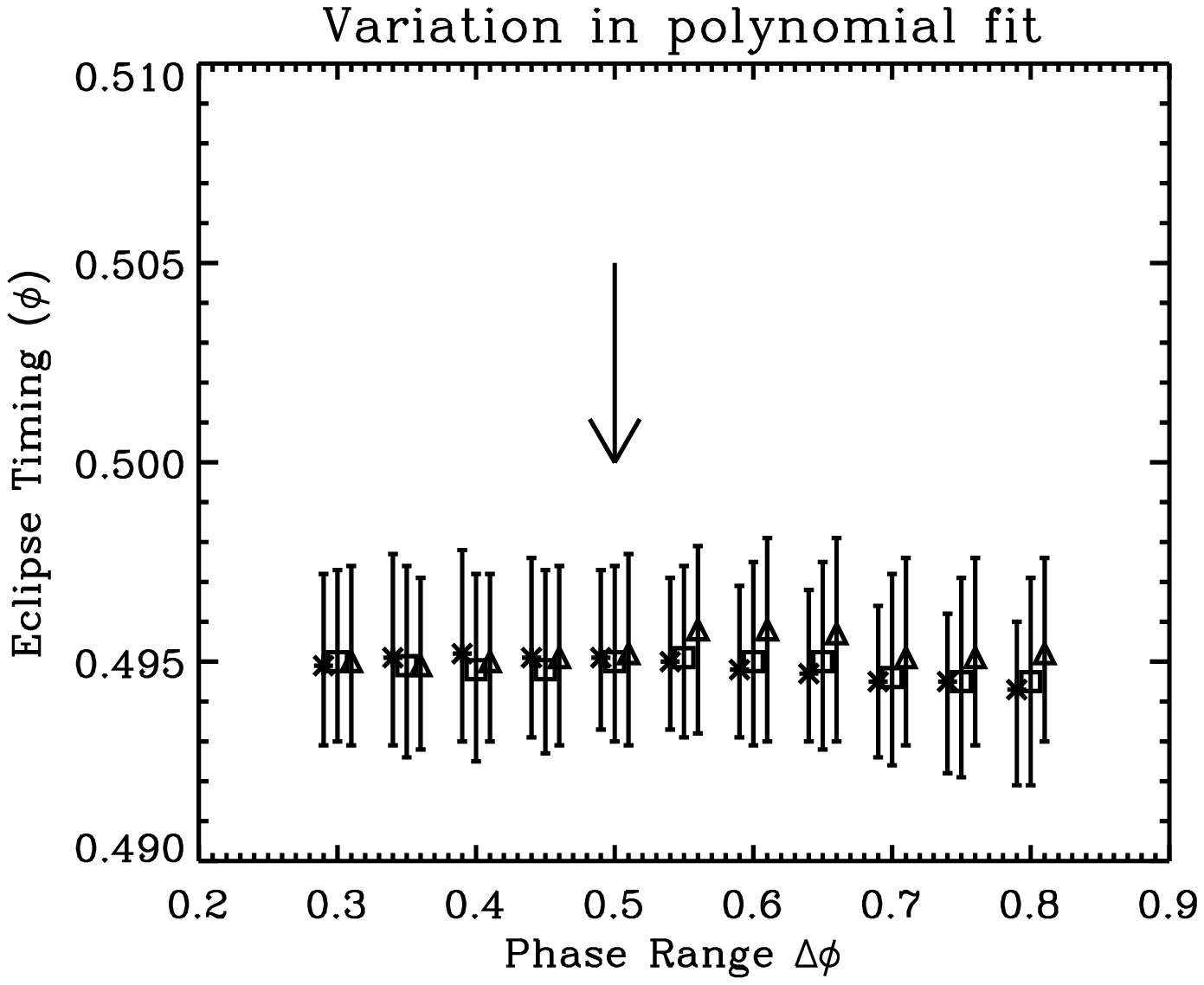,width=0.45\textwidth}
\caption{
The fitted eclipse depth (top panel) and eclipse timing (lower panel) for different choices of how to remove the long-term stellar variability, varying the order of the polynomial (2nd = crosses, 3rd = squares, 4th = triangles), and the total phase-range over which the variability is fit. The arrows indicate the parameters used in our analysis, a phase-range of $ \Delta \phi $=0.5 (0.25$< \phi <$0.75), and a third order polynomial.}
\end{figure}

\begin{figure}
\psfig{figure=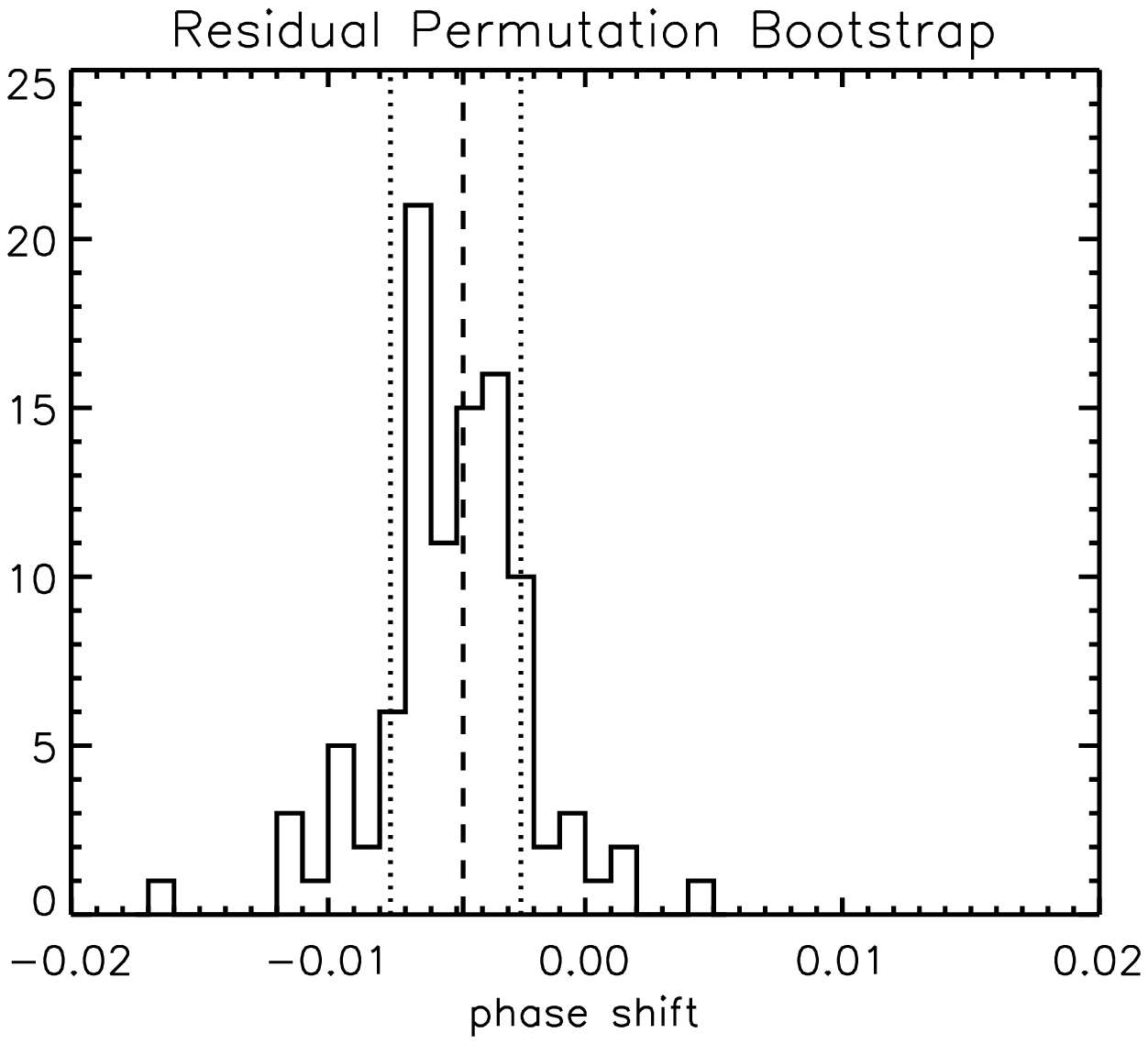,width=0.45\textwidth}
\psfig{figure=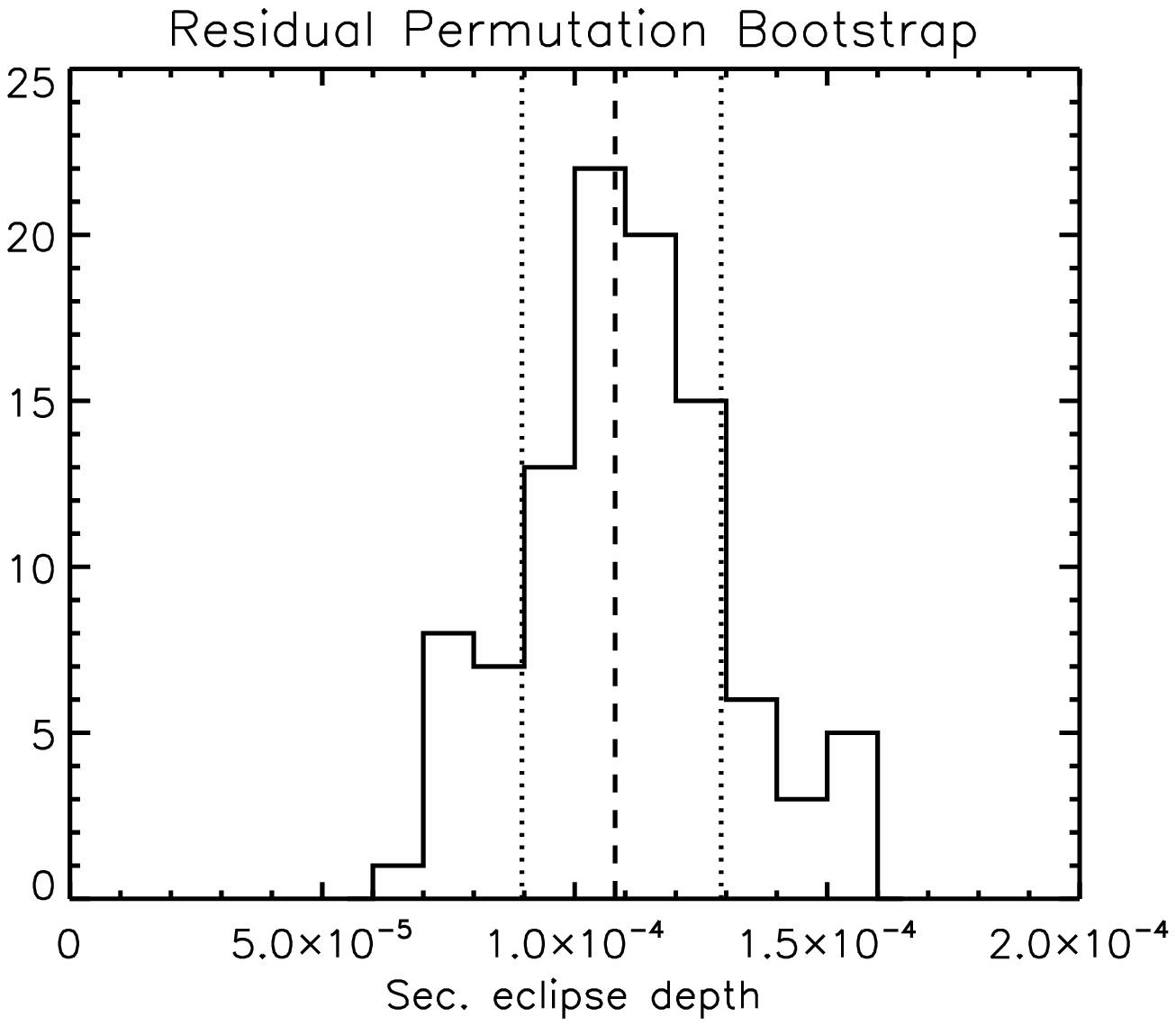,width=0.45\textwidth}
\caption{The distribution of the measured phase shift (top panel) and secondary eclipse depth (lower panel) using the “residual permutation” bootstrap method. The smooth curves are Gaussians fitted to the distribution. With this analysis we find very similar uncertainties in the eclipse timing and depth as from the chi-squared and bootstrap methods.}
\end{figure}

\subsection{Systematic effects}

We further checked for the influence of systematic effects in several ways. First of all related to the correction for the stellar variability, we determined at what level the choice of the fitted phase-range, and the order of the polynomial, influence the results. We varied the range in phase between 0.2$<\phi<$ 0.8 and 0.4$< \phi <$0.6 in steps of 0.025, and the order of the polynomial between 2 and 4. The results are shown in Figure 7. It shows that neither varying the phase-range nor the order of the polynomial used in the fit significantly changes the results. Only when a small phase-range and 4th order polynomial is used, does the eclipse depth deviate towards higher values. We believe that this is due to correlated (instrumental) noise still present at short time-scales, which can be seen as low-amplitude 'wiggles' in the baseline in Figure 5. If the phase-range, over which the polynomials are fitted, is too small, the baseline is not fitted correctly and the resulting eclipse becomes deeper. However, for a phase-range from 0.4 to 0.8, the influence of the instrumental 'wiggles' on the baseline averages out. We therefore believe that a phase-range of 0.5 is well chosen.

The presents of correlated noise could make us underestimate our errors.
We therefore also tried to characterize the uncertainty using the ``residual permutation'' bootstrap method (Winn et al. 2009). The best fitting model was subtracted from the data, after which the residuals were shifted between 1 and 99 points and added back to the model light curve. These light curves were refitted for each individual shift. The resulting distributions of parameter values (Figure 8) represent separate estimates of the uncertainties in the data. With this analysis we find an uncertainty of 1.9$\times$10$^{-5}$ for the eclipse depth and 2.4$\times$10$^{-3}$ for the eclipse timing, hence very similar to what was derived above.  

\subsection{Analysis of the Green, Blue, and White light data}

We also performed a similar analysis for the CoRoT blue channel, green channel, and combined white light curve (blue+green+red channels), as for the red channel data presented in this paper. The resulting phase folded light curves around the secondary eclipse are presented in Figure 9. The noise in the blue and green channels are larger by a factor $\sim$2.5 compared to that for the red channel data, and show no sign of a secondary eclipse.  The white channel light curve has already been presented recently by Alonso et al. (2009) who detect the secondary eclipse at a depth of 6.0$\pm$2.0$\times$10$^{-5}$. We confirm this result at  5.8$\pm 2.0 \times 10^{-5}$, if we use a small phase-range to fit the baseline. However, when we use a larger phase-range, eg. for 0.25$<\phi<$0.75, we measure an eclipse depth of 4.5$\pm$1.5$\times 10^{-5}$, a similar variation as seen in the red light curve, probably due to residual correlated noise. Note that in Alonso et al (2009) the eclipse depth of 6.0$\pm$2.0$\times$10$^{-5}$ is translated to a brightness temperature of 1910$\pm$100 K. However, for the same eclipse depth, with the planet and star parameters as given by Alonso et al. (2008), we calculate that it corresponds to a brightness temperature of T$_{\rm{B}}$=2130$^{+70}_{-100}$ K in the white band. Anyway, since 70\% of the total white light detected by CoRoT originates from the red channel, and the white channel detection is at about a 60\% level of that in the red channel, it suggests that all planet light seen in the secondary eclipse of the white light curve originates (within uncertainties) from the red CoRoT channel.

\section{Comparison with atmosphere models}

\subsection{A simple black body $+$ reflection model}

The measured eclipse depth in CoRoT's red channel corresponds to a planet brightness temperature of 2170$\pm$55 K, assuming a Kurucz model spectrum for the star (with parameters as given in table 1). The error includes the incertainty in the effective temperature of the star. Although it is generally not expected that a hot Jupiter radiates as a black body in this wavelength regime, we can in first instance compare this with the expected range of effective day-side temperatures for CoRoT-2b. If we assume a homogeneous effective temperature over the whole day-side hemisphere, and if we assume the planet to be completely non-reflective at all wavelength (Bond albedo A$_B$=0), then we expect its hemisphere averaged effective day-side temperature to be between 1540 K (for a redistribution factor$ P_n=0.5$, meaning that the planet atmosphere very effectively transports stellar heat from the day-side to the night-side), and T=1830 K (for a redistribution factor P$_n$=0). Assuming the most extreme case, a non-reflective planet with a dynamics free atmosphere where the radiation time scale dominates, the temperatures at the sub-stellar point are significantly higher, boosting the hemisphere-averaged effective day-side temperature to T$_{\rm{max}}$=1965 K. Even this maximum effective temperature is significantly lower than the measured brightness temperature (at 3$\sigma$), which if such planet would radiate as a black body, would produce a secondary eclipse with a depth of 4$\times$10$^{-5}$ in CoRoT's red channel. 

In addition to the optical detection presented here, 
CoRoT-2b has recently also been observed in the infrared using the Spitzer Space Telescope, showing eclipse depths of 5.10$\pm 0.42 \times 10^{-3}$ and 4.1$\pm 1.1 \times 10^{-3}$ at 4.5 and 8 micron respectively (Gillon et al. 2009). These measurements correspond to brightness temperatures of 
T=1805$\pm$70 K and T=1325$\pm$180 K at 4.5 and 8.0 micron respectively, significantly lower than in the optical.

Keeping the important caveats in mind, that the spectra of neither self-luminous gaseous planets (and brown dwarfs) nor highly irradiated planets are expected to closely resemble Planck curves, we show in Figure \ref{TA} the planet's day-side temperature versus the albedo in the red channel for a zeroth-order model, consisting of a thermal black body spectrum plus a constant geometric albedo over the red channel pass-band. The grey band indicates those values allowed by the data within their 1$\sigma$ uncertainties, implying a geometric albedo in the range 0.06$<$Ag$<$0.2.

\begin{figure*}

\hspace{1cm}
\hbox{\psfig{figure=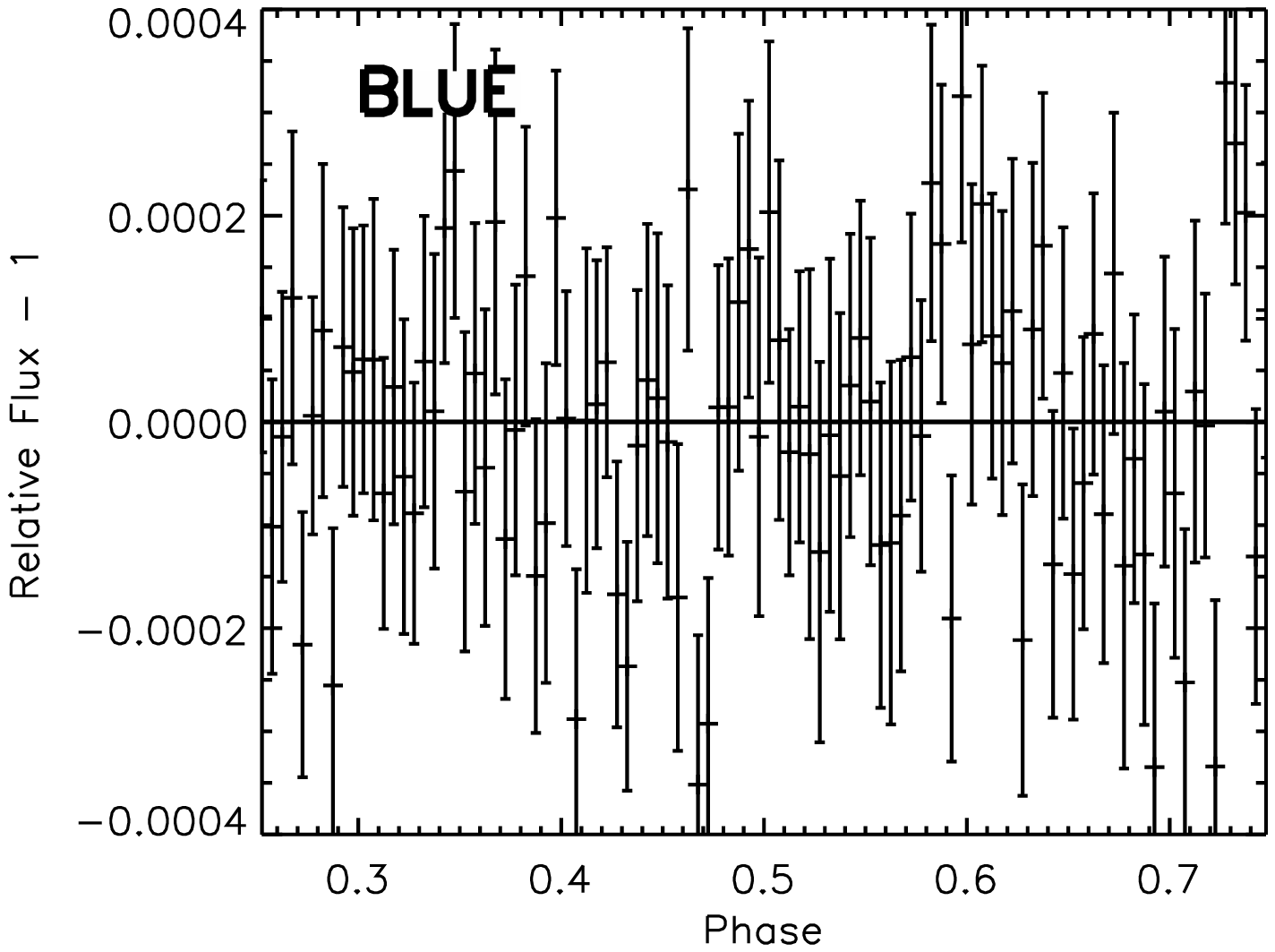,width=0.4\textwidth}
\psfig{figure=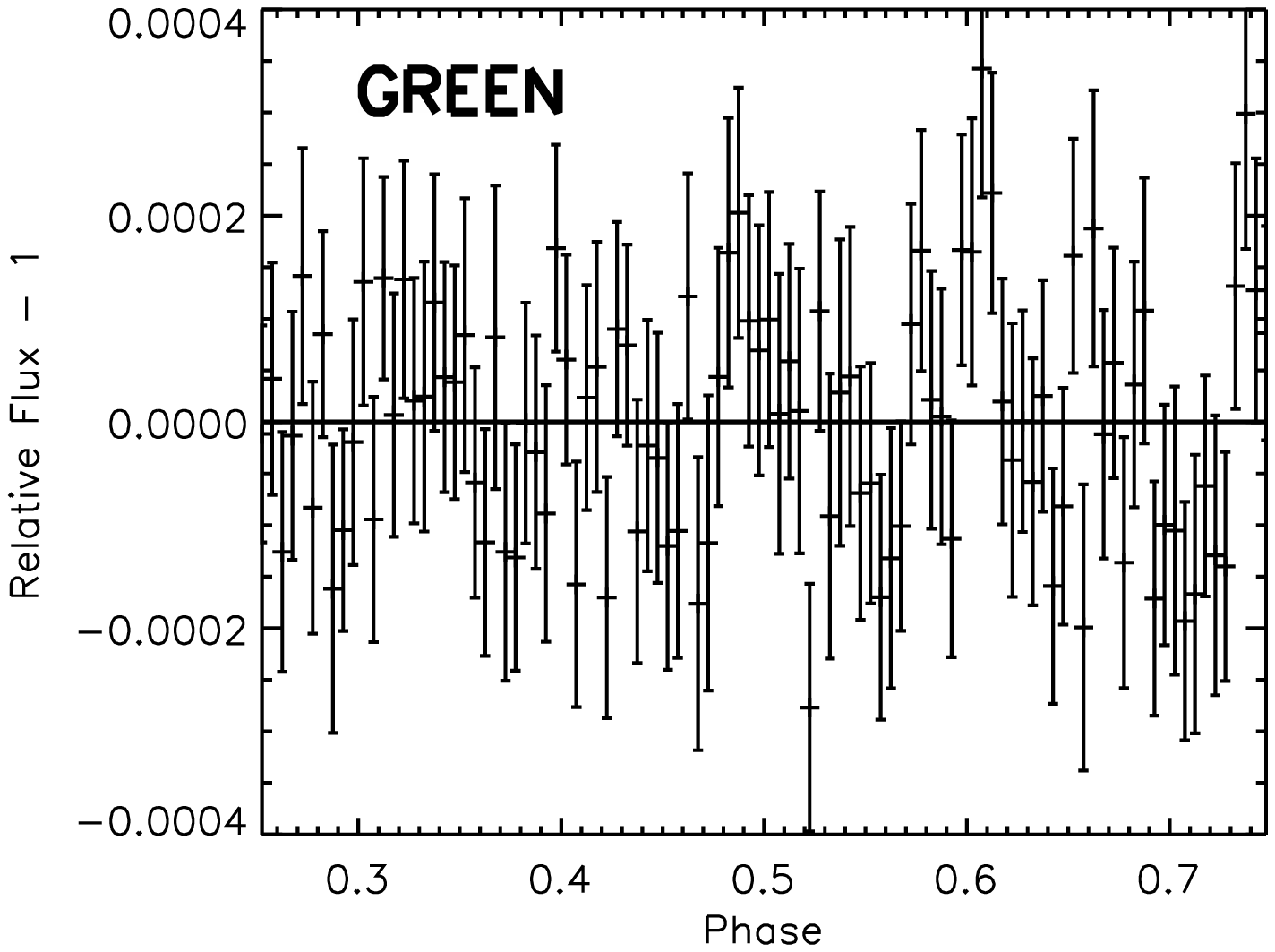,width=0.4\textwidth}}

\hspace{1cm}
\hbox{
\psfig{figure=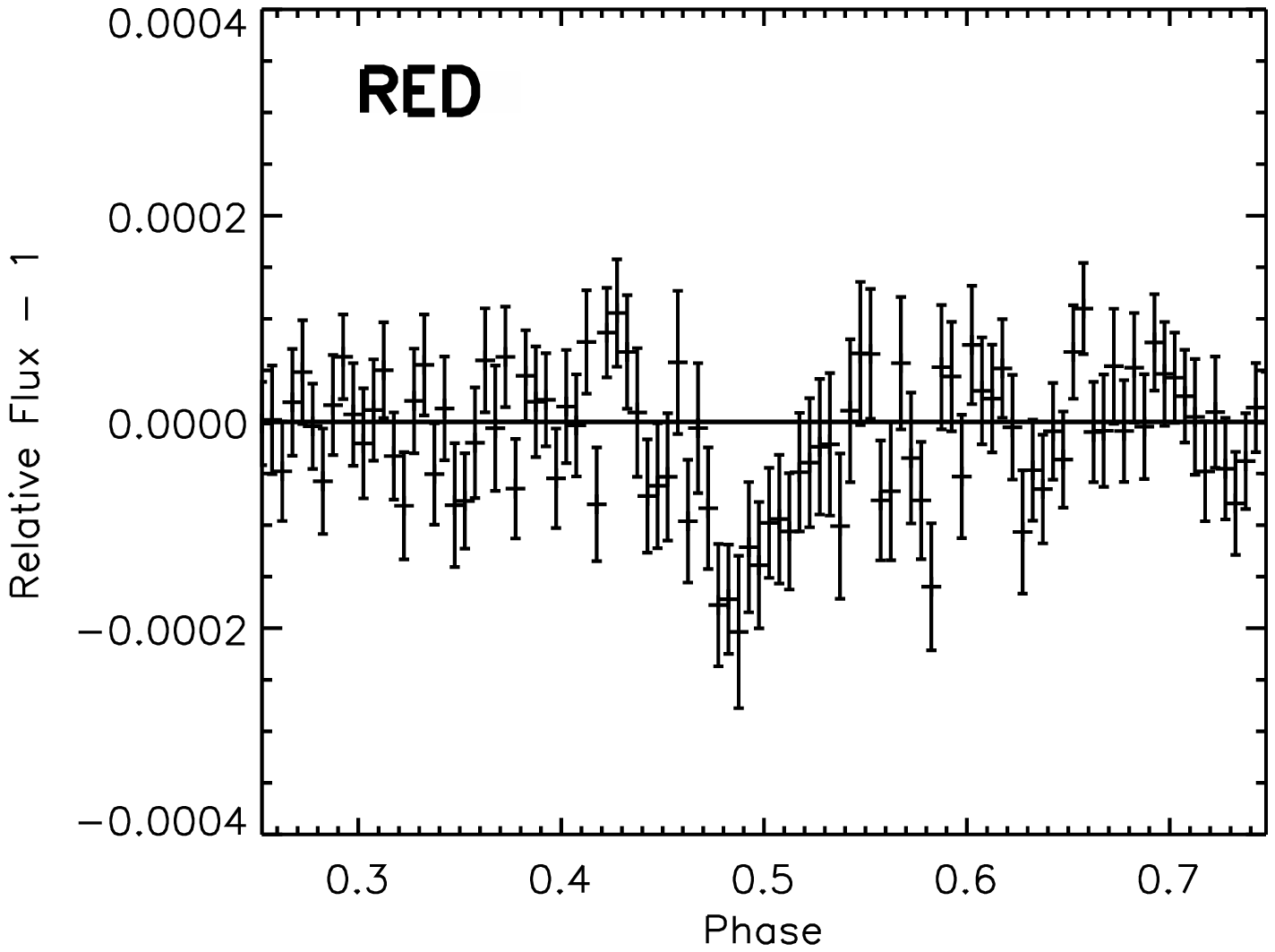,width=0.4\textwidth}
\psfig{figure=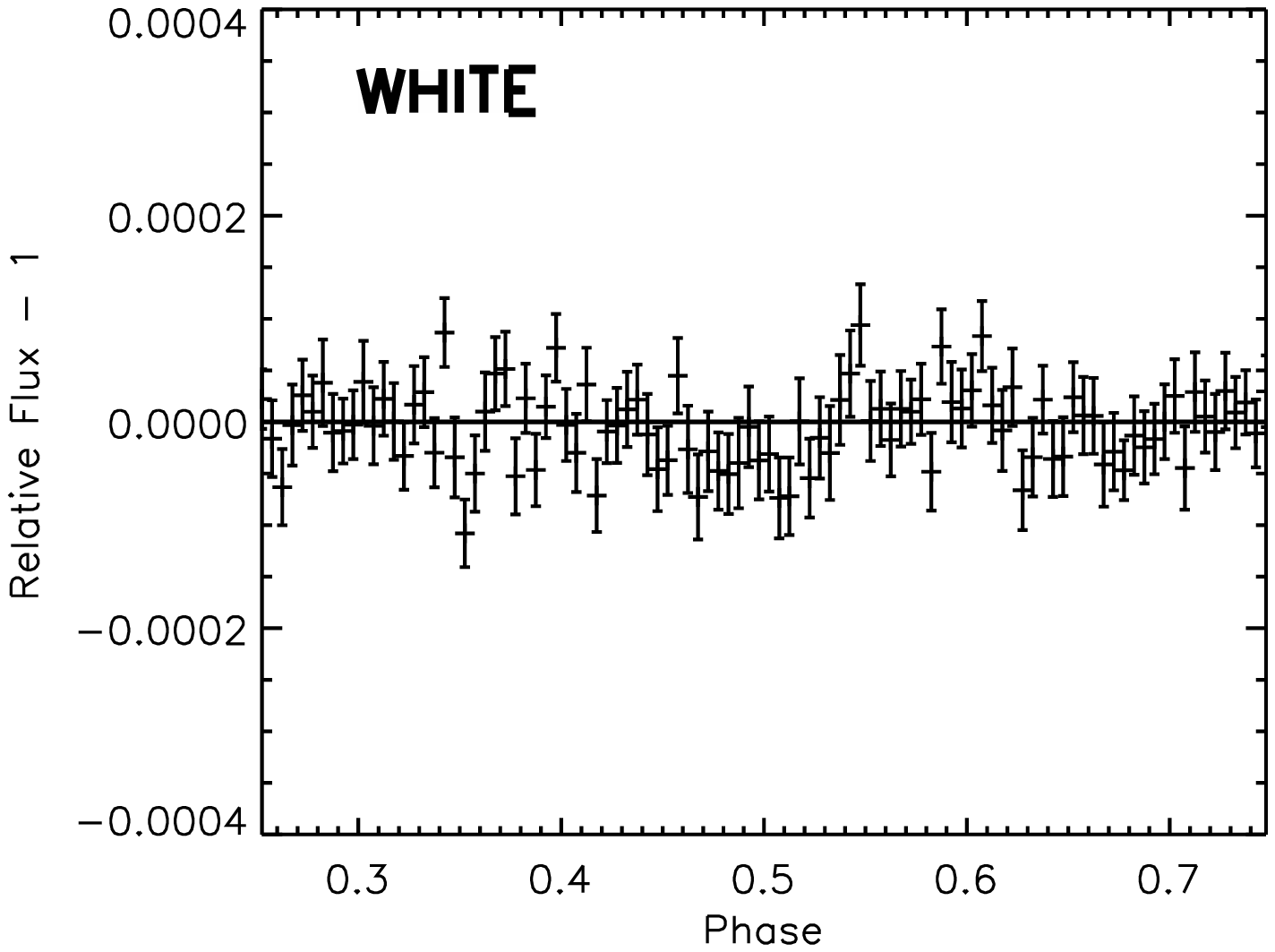,width=0.4\textwidth}}
\caption{The phase-folded light curves for the blue channel, green channel, red channel, and combined white data. Although the secondary eclipse in the white light curve data is formally detected at a 3$\sigma$ level, at least in our analysis the fitted eclipse depth varies significantly depending on the range in phase over which the stellar variability is fitted and removed.} 
\end{figure*}

\begin{figure}
\psfig{figure=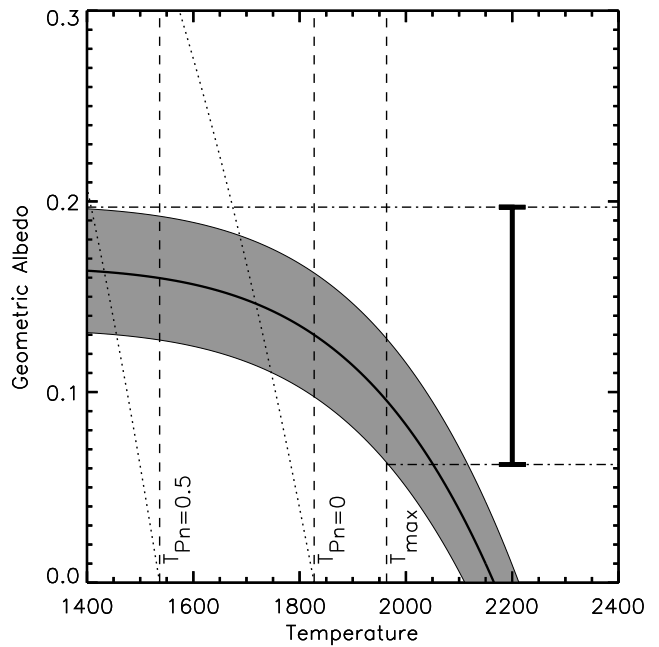,width=0.5\textwidth}
\caption{\label{TA} A simple parametrization of the optical flux from CoRoT-2b. The planet's uniform day-side temperature is shown versus the albedo in the red channel, for a zeroth-order model consisting of a thermal black body spectrum plus a constant geometric albedo over the red channel pass-band. The three vertical dashed lines show the day-side effective temperatures, for a non-reflective planet with no redistribution of stellar energy towards the night-side (T$_{\rm{Pn=0}}$ = 1830 K), for a non-reflective planet with half of the stellar energy transported to the night-side (T$_{\rm{Pn=0.5}}$ = 1540 K), and for a non-reflective dynamics free atmosphere (T$_{\rm{max}}$=1965 K). The dotted lines show the same as the dashed lines, but assuming a wavelength-independent Lambert scattering, with the Bond albedo, A$_B$ = 3/2 A$_g$. The grey band covers those values allowed by the data within their 1$\sigma$ uncertainties, indicating a geometric albedo of 0.06$<$Ag$<$0.2 for this simple model. However, note that the spectra of neither self-luminous gaseous planets nor highly irradiated planets are expected to closely resemble Planck curves, thus this in itself is no proof for reflected light.}
\end{figure}

\subsection{Burrows, Ibgui \& Hubeny atmosphere code}
\begin{figure*}
\psfig{figure=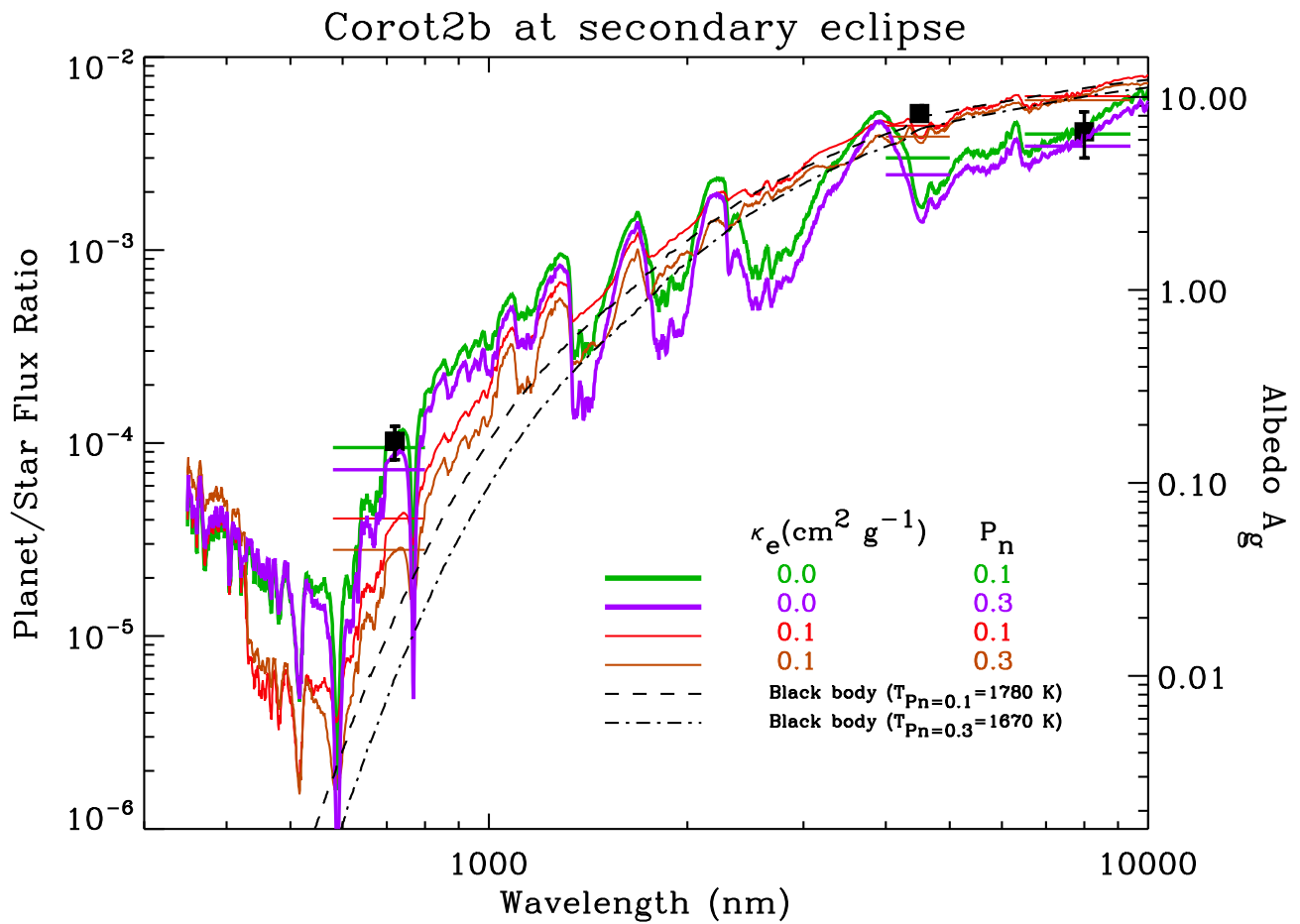,width=\textwidth}
\caption{\label{adam} The ratio of planet to star flux (left axis) and geometric albedo (right axis) are shown as function of wavelength, calculated using a self-consistent atmosphere code (Burrows et al. 2008). The green and magenta lines indicate model spectra for stellar heat redistribution factors of P$_n$=0.1 and P$_n$=0.3, respectively. The red and brown lines are for the same redistribution factors, but now with an extra absorber distributed high up in the atmosphere. Superimposed on the model spectra are the CoRoT measurement and Spitzer measurements of Gillon et al. (2009)}. The horizontal bars show the expected model-fluxes for the CoRoT and Spitzer pass-bands. To guide the eye, but keeping in mind their limited physical meaning, the flux ratios for a hypothetical planet with a black body spectrum for an effective temperature of 1780 K (P$_n$=0.1) and 1670 K (P$_n$=0.3), are indicated by the dashed and dot-dashed lines respectively. 
\end{figure*}

We model the ratio of planet to star flux as function of wavelength, using the self-consistent atmosphere code and solution techniques (Hubeny \& Lanz 1995; Burrows, Hubeny \& Sudarsky 2005; Burrows et al. 2007) as employed for HD209458b (Burrows, Ibgui \& Hubeny 2008). We produced four sets of spectra, which are shown in Figure \ref{adam} as planet/star flux ratios (left axis) and planet albedos A$_g$ (right axis) as function of wavelength. All models are calculated for an atmosphere with a solar metallicity. The green and magenta lines indicate the spectra for stellar heat redistribution factors of P$_n$ =0.1 and P$_n$=0.3 respectively. The red and brown lines are for the same redistribution factors, but now with an extra absorber distributed everywhere in the atmosphere at pressures lower than ~0.014 bars, with an opacity of $\kappa$=0.01 cm$^2$g$^{-1}$ in the wavelength range 0.43 $\mu$m$<$$\lambda$$<$1.0 $\mu$m. Such an 'extra absorber' in the optical has been invoked to explain the photometric reversals and flux enhancements possibly caused by a thermal inversion as seen in several of the highest irradiated planets, including HD209458b, which may also be present in CoRoT-2b.

The CoRoT and Spitzer (Gillon et al. 2009) measurements are superimposed on the model spectra. 
The horizontal bars show the expected model-fluxes for these observations, properly integrated over the pass-bands. The lengths of the bars indicate the effective width of this band. Note that for all models, A$_g$ exceeds unity long-ward of $\sim$1 $\mu$m. Since in this context A$_g$ is strictly defined as the ratio of planet to stellar flux at the same wavelength, this does not violate energy conservations, because in the infrared the planet flux is dominated by absorption and re-emission of the star's optical flux. To guide the eye, but keeping in mind their limited physical meaning, the flux ratios for a hypothetical planet with a black body spectrum for an effective temperature of 1780 K (P$_n$=0.1) and 1670 K (P$_n$=0.3), are indicated by the black dashed and dot-dashed lines respectively. 

In the models without an extra optical absorber, broad absorption bands are present in the infrared, mainly from water, methane, and CO, and relatively narrow absorption features can be seen in the optical from sodium and potassium. The flux ratios strongly decrease from infrared to optical wavelengths until Rayleigh scattering off H$_2$ and He becomes dominant at $<$600 nm. The P$_n$=0.1 model spectrum provides a good fit to the CoRoT measurement, giving a planet/star flux ratio of 9.5$\times$10$^{-5}$ in CoRoT's red channel, corresponding to geometric albedo of 0.152.  The models that include an extra absorber provide only 30-40\% of the flux seen in the CoRoT red channel.

\section{Discussion and conclusions}

The measured brightness temperature of 2170$\pm$50 K in CoRoT's red channel is significantly higher than the maximum possible hemisphere-averaged effective day-side temperature for this planet. However, as we have stressed before, it is not expected that such planet would radiate as a black body in this wavelength regime, and therefore does not directly imply that the light is dominated by reflected star light. The measured flux corresponds to a geometric albedo of A$_g$=0.164$\pm$0.032, with A$_g$ strictly defined as above. 

We compared the optical secondary eclipse depth, and Spitzer measurements at 4.5 and 8 $\mu$m (Gillon et al. 2009) with the Burrows, Ibgui \& Hubeny (2009) atmosphere models for CoRoT-2b, assuming a range of stellar energy redistribution factors, and assuming the presence or not of a strong optical absorber high up in the planet's atmosphere. It is currently thought that the presence or absence of such strong absorbers in hot Jupiter atmospheres is connected to the level of irradiation (Burrows et al. 2005; Fortney et al. 2008). For planets that receive less stellar flux, this absorber condenses out of the upper atmosphere, e.g. for HD189733b, while at higher irradiation levels this absorber would cause a thermal inversion layer, as thought to be the case for HD209458b. It is often quoted that the strong optical absorbers TiO and VO are probably the responsible species (Fortney et al. 2008). However, this has recently been questioned. Using a radiative-convective radiative-transfer model and a model of particle settling in the presence of turbulent and molecular diffusion, it is found that it is unlikely that VO plays a role in producing an upper atmosphere thermal inversion (Spiegel, Silverio, \& Burrows 2009). Furthermore, the models imply that a heavy species such as TiO can only persist in a planet's upper atmosphere with a strong macroscopic mixing process (Spiegel et al. 2009). 

Purely from its high level of irradiation, in comparison with other planets, it would be expected that CoRoT-2b exhibits a temperature inversion, implying that the unspecified extra absorber is not condensed out. However, the optical eclipse depth is only fitted well by a model atmosphere without such extra absorber, while the models with this extra absorber underestimate the optical flux by more than a factor two. This favours models for CoRoT-2b without a significant thermal inversion layer.

We can estimate the fraction of reflected light in CoRoT's red channel for the set of models from the spectral behaviour at $<$600 nm, where the planet flux {\sl is} dominated by reflected star light. It indicates that for those models without an additional high-altitude absorber, the reflected light component should contribute at a level of 1$-$2$\times10^{-5}$, meaning that reflected light forms 10$-$20\% of the planet flux is CoRoT's red channel. For those models with an extra absorber, the reflected light component is significantly lower, at a level of $\sim$5$\times 10^{-6}$ in CoRoT's red channel. 

\begin{figure}
\psfig{figure=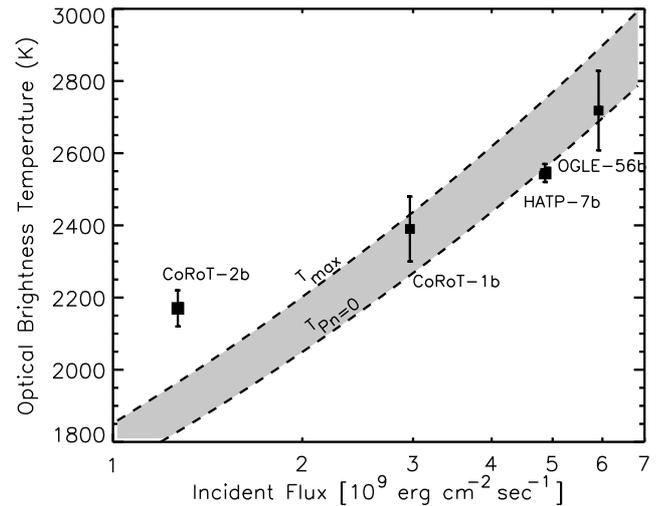,width=0.5\textwidth}
\caption{\label{comp} The observed optical day-side brightness temperature, as determined from secondary eclipse observations, as function of the incident star flux, for the four planets detected so far. The grey area indicates brightness temperatures between T$_{\rm{max}}$ and T$_{{\rm P_n=0}}$ (but do note the limited physical meaning of comparing brightness temperatures with effective temperatures). It shows that CoRoT-2b is not only the coolest planet of the four, it is also the only planet for which the optical brightness temperature exceeds T$_{\rm{max}}$. Although we argue that reflected light plays a small role, we believe this is mainly caused by CoRoT-2b having a different atmospheric temperature structure than the other planets, possibly due to the absence of a inversion layer.}
\end{figure}

So far, optical secondary eclipses have been detected for three other hot Jupiters, OGLE-TR-56b (Sing \& Lopez-Morales 2009), CoRoT-1b (Snellen et al. 2009; Alonso et al. 2009), and HATP-7b (Borucki et al. 2009). It is very interesting to compare the measured brightness temperatures with the maximum possible effective day-side temperatures for these planets, as shown in Figure \ref{comp}. The grey band indicates the area for which brightness temperatures fall between the maximum hemisphere-averaged effective dayside temperature T$_{\rm{max}}$ and T$_{{\rm P_n=0}}$. It shows that CoRoT-2b is not only the coolest planet of the four, it is also the only planet for which the optical brightness temperature exceeds T$_{\rm{max}}$. We mention again that it is not expected that the spectra of irradiated hot Jupiters follow a Planck curve, and that therefore the direct comparison of brightness temperature with the planet's effective temperature has limited physical meaning. However, it is an indication that the atmospheric structure of CoRoT-2b is different from the other planets, which we believe could be due to the absence of a thermal inversion layer in CoRoT-2b. Indeed, Figure \ref{adam} shows that the optical spectra of planets with an extra absorber at high altitude, are relatively suppressed in flux compared to those without such absorber. 
This is because in the models with a weak extra absorber in the atmosphere,
as a result of such absorption the optical flux can be lower than for models
without such absorber (Burrows et al. 2008). As one first raises the amount of extra optical absorber at low pressures (high altitudes), the temperature in the upper atmosphere (at the same low pressures) goes up. This is the source of the extra mid-infrared flux. At the same time, it absorbs in the optical thereby suppressing emergent optical flux. Furthermore the temperature at higher pressures goes down, which is the origin of some of the near-infrared and also emergent optical flux.  As one raises the amount of extra absorber further, for large stellar fluxes at the sub-stellar point, the temperature at low pressure continues to go up, as does the mid-infrared flux. Now, the temperature in the near-infrared photospheric regions also starts to rise, which at this point also leads to an increase again of the emergent optical flux. Hence, in this regime the optical flux is not  a monotonic function of the temperature at low pressure, but first goes down and then up (Burrows et al. 2008).

\section*{Acknowledgments}

We thank the CoRoT team for making the CoRoT data publicly available in a high quality and comprehensible way, which forms the basis of this study. The CoRoT space mission, launched on Dec 27th 2006, was developed and is operated by the CNES, with participation of the science program of ESA, ESTEC/RSSD, Austria, Belgium, Brasil, Germany, and Spain. A.B. acknowledges support under NASA grant NNX07AG80G and under JPL/Spitzer Agreements 1328092, 1348668, and 1312647.

\end{document}